\documentclass[12pt]{article}
\usepackage{epsfig}
\begin{document}
\hspace*{13cm}IU-MSTP/50 \\
\hspace*{13cm}hep-lat/0206014 \\
\hspace*{13.9cm}June, 2002
\begin{center}
 {\Large\bf Spectrum of the Hermitian Wilson-Dirac 
Operator for a Uniform Magnetic Field in Two Dimensions}
\end{center}

\vspace*{1cm}
\def\thefootnote{\fnsymbol{footnote}}
\begin{center}{{\sc Hiroshi Kurokawa}$^1$ and {\sc Takanori Fujiwara}$^2$} 
\end{center}
\vspace*{0.2cm}
\begin{center}
$^1${\it Graduate School of Science and Engineering, Ibaraki University, 
Mito 310-8512, Japan}

$^2${\it Department of Mathematical Sciences, Ibaraki University, 
Mito 310-8512, Japan} 
\end{center}
\vfill
\begin{center}
{\large\sc Abstract}
\end{center}
\noindent
It is shown that the eigenvalue problem for the hermitian 
Wilson-Dirac operator of for a uniform magnetic field in two 
dimensions can be reduced to one-dimensional problem described 
by a relativistic analog of the Harper equation. An explicit 
formula for the secular equations is given in term of a set of 
polynomials. The spectrum exhibits a fractal structure in the 
infinite volume limit. An exact result concerning the index 
theorem for the overlap Dirac operator is obtained. 


\eject
The effects of magnetic fields on two-dimensional systems 
have attracted continual interests from condensed matter 
to elementary particle physics. Magnetic fields not only
bundle energy spectrum but also rearrange the eigenvalues 
as they continuously changes, giving rise to rich band structure. 
The most impressive is the energy spectrum 
of tightly bounded electron in a uniform magnetic field. It is 
well-known that the energy spectrum of this system has a fractal 
structure known as the butterfly diagram \cite{hof}. The key 
ingredient that underlies such variety is the topological nature 
of magnetic fields. 

Relativistic analogue of the Hamiltonian of tightly bounded electron 
is the hermitian Wilson-Dirac operator (HWDO), which attracts renewed 
interests in the context of overlap Dirac operator (ODO) describing chiral 
gauge theories on the lattice \cite{neub1}. Magnetic field also 
affects the 
spectrum of HWDO, giving rise to interesting phenomena such as chiral 
anomaly \cite{HLN}. We also expect the spectrum has a fractal structure 
similar to the butterfly diagram. 

In this paper we investigate the spectrum of HWDO for a uniform magnetic 
field in two dimensions. We show that the two dimensional system can be 
converted to one dimensional problem for an arbitrary uniform magnetic 
field and, as the consequence, the spectrum can be characterized by a 
set of polynomials, which enables us not only to understand the fractal 
structure of the spectrum but also to compute the exact index of the ODO. 

The spectral flow of HWDO for the one-parameter family of link variables
was investigated in Ref. \cite{NN2} and the connection to chiral anomaly 
was elucidated. Recently, one of the present author reanalyzed the same 
system \cite{fuji} and gave some exact results on the spectrum for 
a particular set of uniform magnetic fields, for which the index 
of the ODO can be obtained rigorously. The purpose of this paper is 
to extend them for an arbitrary uniform magnetic field. 

We consider a two-dimensional square lattice of unit lattice 
spacing and of sides $L$. The HWDO $H$ is defined by 
\begin{eqnarray}
  \label{eq:hWDop}
  H\psi(x)=\sigma_{3}\Biggl\{(2-m)\psi(x)-\sum_{\mu=1,2}
  \Biggl(\frac{1-\sigma_\mu}{2}U_\mu(x)\psi(x+\hat\mu)
  +\frac{1+\sigma_\mu}{2}U^\ast_\mu(x-\hat\mu)\psi(x-\hat\mu)
  \Biggr)\Biggr\}~, 
\end{eqnarray}
where $\sigma_\mu$ ($\mu=1,2,3$) are the Pauli matrices and 
$\hat\mu$ stands for the unit vector along the $\mu$-th axis. 
The link variables $U_\mu(x)$ and the lattice fermion field 
$\psi(x)$ are subject to periodic boundary conditions. 
In the context of ODO $m$ must be chosen to satisfy $0<m<2$ in 
order for the correct continuum limit to be achieved. We adopt 
$m=1$ unless otherwise specified.

We are interested in the spectrum of $H$ for the link variables 
of the form 
\begin{eqnarray}
  \label{eq:lv}
  U_1(x)=\exp\Biggl[-it\frac{2\pi}{L}\overline x_2
  \delta_{\overline x_1,L-1}\Biggr]~, \qquad
  U_2(x)=\exp\Biggl[it\frac{2\pi}{L^2}\overline x_1\Biggr]~,
\end{eqnarray}
where $t$ is an arbitrary parameter and $\overline x_\mu$ stands for 
the periodic lattice coordinates defined by $\overline x_\mu=x_\mu$ 
for $0\leq x_\mu<L$ and $\overline {x_\mu+L}=\overline x_\mu$. 
For $t=Q$ being an integer, the magnetic field 
$F_{12}(x)=-iU_1(x)U_2(x+\hat1)U_1^\ast(x+\hat2)U_2^\ast(x)$ 
becomes a constant and the flux per plaquette normalized by $2\pi$ 
is given by 
\begin{eqnarray}
  \label{eq:fs}
  \alpha\equiv\frac{1}{2\pi}F_{12}(x)=\frac{Q}{L^2}~.
\end{eqnarray}
Hence $Q$ is nothing but the topological charge \cite{lus1}. 
The parameter $t$ connects continuously the link variables with 
various uniform magnetic fields belonging to different topological 
sectors classified by the integer topological charge 
\cite{lus1,NN2,fuji}. 

The characteristics of the spectrum of HWDO found in \cite{fuji} can 
be summarized as; (1) The eigenvalues at integral values of $t$ are 
separated by several gaps and form clusters. (2) For noninteger $t$ the 
$H$ has in general $2L^2$ distinct eigenvalues and rearrangements 
of eigenvalues between the clusters occur in a characteristic way as 
$t$ increases continuously from an integer to the next integer.
By carefully inspecting the spectral flows one realizes that for 
special integer values of $t=r$ with $r$ being an arbitrary divisor 
of $L$ each eigenvalue is exactly $r$-ply degenerate. This can be 
proven by noting that $H$ can be block diagonalized into $r$ 
$2sL\times 2sL$ matrices each describing a one-dimensional lattice 
system of degrees of freedom $2sL$ \cite{fuji}. In what follows we 
will generalize the argument of Ref. \cite{fuji} concerning the 
characterization of the spectrum of $H$ for special values $t=r$ 
or $t=L^2/r$ to arbitrary integer values of $t$. In doing this it 
is helpful to note the following general facts; (1) The eigenvalues 
$\lambda$ of $H$ are bounded by $|\lambda|\leq|2-m|+2$ \cite{neub2}. 
(2) The spectrum of $H$ is periodic in $t$ with a period $L^2$. (3) The 
spectrum is an odd functions in $t$. Hence, it suffices
to analyze the eigenvalue spectrum for $0\leq t\leq L^2/2$. 

We first analyze the spectrum for $t$ being an integer multiple of $L$. 
Let $L$ be a product of two positive integers $r$ and $s$. Then $t$ 
can be expressed as $t=nL^2/r=nsL$, where $n$ is some integer coprime 
with $r$. Then we have $U_1(x)=1$ and 
$U_2(x)=\exp[2\pi in\overline x_1/r]$. Since 
the link variable is independent of $x_2$ and periodic in $x_1$ with a 
period $r$, we can simplify the eigenvalue problem for $H$ by the 
following Fourier transformation 
\begin{eqnarray}
  \label{eq:ft}
  \varphi(y;p,q)
  =\frac{1}{sL}\sum_{l=0}^{s-1}\sum_{x_2=0}^{L-1} e^{-iql-ipx_2}
  \psi(rl+y,x_2)~,
\end{eqnarray}
where $y$ ranges between $0$ and $r-1$. The Fourier momenta $p$ and 
$q$ are given by
\begin{eqnarray}
  \label{eq:fm}
  p=\frac{2\pi}{L}k~, \quad q=\frac{2\pi}{s}j~. \quad
  (k=0,1,\cdots,L-1~,~~
  j=0,1,\cdots, s-1)
\end{eqnarray}
$H$ is then block-diagonalized into $sL$ $2r\times2r$ hermitian 
matrices $h(p,q)$ given by
\begin{eqnarray}
  \label{eq:h}
  h(p,q)=\pmatrix{B(p,q) & C(p,q) \cr C^\dagger(p,q) & -B(p,q)}~,
\end{eqnarray}
where the first (second) row acts on the upper (lower) component of 
$\varphi(y;p,q)$.  $B(p,q)$ and $C(p,q)$ are defined by
\begin{eqnarray}
  \label{eq:BC}
  (B(p,q))_{y,y'}&=&-\frac{1}{2}\delta^{(q)}_{y+1,y'}
  +\Biggl\{1-\cos\Biggl(p+\frac{2\pi ny}{r}\Biggr)\Biggr\}
  \delta^{(q)}_{y,y'}
  -\frac{1}{2}\delta^{(q)}_{y,y'+1}~, \nonumber \\
  (C(p,q))_{y,y'}&=&\frac{1}{2}\delta^{(q)}_{y+1,y'} 
  +\sin\Biggl(p+\frac{2\pi ny}{r}\Biggr)\delta^{(q)}_{y,y'}
  -\frac{1}{2}\delta^{(q)}_{y,y'+1}~.
\end{eqnarray}
The $\delta^{(q)}_{y,y'}$ is the Kronecker $\delta$-symbol
for $0\leq y,y'\leq r-1$ and satisfies the twisted boundary conditions
\begin{eqnarray}
  \label{eq:tbc}
  \delta^{(q)}_{y,r}=e^{-iq}\delta_{y,0}~, \qquad
  \delta^{(q)}_{r,y}=e^{iq}\delta_{y,0}~.
\end{eqnarray}
We thus find that the original two-dimensional system of degrees of 
freedom $2L^2$ is decomposed into $sL$ one-dimensional systems of 
degrees of freedom $2r$, each described by $h(p,q)$. The corresponding 
eigenvalue problem is just a relativistic analog of the Harper 
equation \cite{Harp}. 

The eigenvalues of $H$ are determined by the secular equation 
$\det(h(p,q)-\lambda)=0$. It takes the form 
\begin{eqnarray}
  \label{eq:seq}
  \det(h(p,q)-\lambda)=f^{(n)}_r(\lambda;p)
  -\frac{(-1)^{r-1}}{2^{r-4}}\sin^2\frac{rp}{2}
  \sin^2\frac{q}{2}=0~,
\end{eqnarray}
where $f^{(n)}_r(\lambda;p)=\lambda^{2r}+\cdots$ is a polynomial of 
order $2r$ and is defined by 
\begin{eqnarray}
  \label{eq:frnlp}
  f^{(n)}_r(\lambda;p)=\det(h(p,0)-\lambda)~.
\end{eqnarray}
The $q$ dependent term in (\ref{eq:seq}) can be easily found from 
the the explicit form of $h(p,q)$. For $n$ and $r$ being coprime 
$h(p,0)$ and $h(0,0)$ are related by an orthogonal transformation 
and $f^{(n)}_r(\lambda;p)$ then becomes independent of $p$. We simply 
write it as $f^{(n)}_r(\lambda)$. Explicit forms of $f_r^{(1)}(\lambda)$ 
are given in Ref. \cite{fuji} for $r=1,\cdots,6$. 

Later we need to consider (\ref{eq:frnlp}) for $n$ and $r$ not 
necessarily being coprime. In general $f^{(n)}_r(\lambda;p)$ 
depends on $p$ and satisfies the factorization property
\begin{eqnarray}
  \label{eq:fp}
  f_r^{(n)}(\lambda;p)=\prod_{j=0}^{s'-1}
  \Biggl(f_{r'}^{(n')}(\lambda)-\frac{(-1)^{r'-1}}{2^{r'-4}}
  \sin^2\frac{r'p}{2}\sin^2\frac{\pi j}{s'}\Biggr) ~,
\end{eqnarray}
where $n'=n/s'$ and $r'=r/s'$ with $s'$ being the greatest common divisor 
of $n$ and $r$. The easiest way to show this is to consider the case 
of $r=L$ and $t=nL^2/r=n'L^2/r'=n's'L$. Since $s=1$, the only allowed 
value of $q$ is $0$. The secular equation (\ref{eq:seq}) then becomes 
$f_r^{(n)}(\lambda;p)=0$. On the other hand the same set of eigenvalues 
must be reproduced by the expressions (\ref{eq:seq}) with the 
substitution $r,n\rightarrow r',n'$ and $q \rightarrow 2\pi j/s'$ 
for $j=0,\cdots,s'-1$. The identity (\ref{eq:fp}) then follows from 
these.  

We have shown that the eigenvalues of $H$ for $t$ being any integer 
multiple of $L$ are completely characterized by a set of functions 
$f_r^{(n)}(\lambda)$. 
We now extend the results to an arbitrary integer $t$. Let $n$ and $s$ be 
positive coprime integers such that $t/L=n/s$, then we may find a positive 
integer $r$ satisfying $t=nr$ and $L=rs$. Denoting $x_2$ by two nonnegative 
integers $y$ and $l$  ($0\leq y<s,~ 0\leq l<r$) as $x_2=sl+y$, we see 
that the link variables (\ref{eq:lv}) are independent of $l$ or, 
equivalently, periodic in $y$ with a period $s$. We then define new 
variables $\Psi(z;q)$ by 
\begin{eqnarray}
  \label{eq:vphi2}
  \Psi(z;q)=\frac{1}{L}\sum_{y=0}^{s-1}\sum_{l=0}^{r-1}
  e^{-i(py+ql)}\psi(x_1,sl+y)~, 
\end{eqnarray}
where $q$, $p$ and $z$ are defined by 
\begin{eqnarray}
  \label{eq:qp}
  q=\frac{2\pi j}{r}~, \quad
  p=\frac{2\pi nk}{s}+\frac{2\pi j}{L} ~, \quad
  z=kL+\overline{x}_1~. \quad (j=0,\cdots,r-1,~k=0,\cdots,s-1)
\end{eqnarray}
Since the shift $z \rightarrow z+sL$ simply corresponds 
to $k \rightarrow k+s$ or, equivalently, to $p \rightarrow p+2\pi n$, 
we may regard $\Psi(z;q)$ as functions of $z$ with a period $sL$. 
Here a miracle occurs. In the new variables $\Psi(z;q)$ defined on the 
one-dimensional periodic lattice, $H$ is again block diagonalized into 
$r$ hermitian $2sL\times 2sL$ matrices of the form (\ref{eq:h}) and 
(\ref{eq:BC}).  The concrete expressions of $B$ and $C$ are obtained 
from (\ref{eq:BC}) by making the following substitutions 
\begin{eqnarray}
  \label{eq:sLcase}
  y,~r,~p,~q\quad
  \rightarrow \quad z,~sL,~\frac{q}{s},~0~.
\end{eqnarray}
The secular equations for the eigenvalues of $H$ at $t=nr$ are then 
given by $f_{sL}^{(n)}(\lambda;q/s)=0$. Noting the factorization formula 
(\ref{eq:fp}), we find that all the eigenvalues of $H$ at $t=nr$ are 
determined by the following set of secular equations
\begin{eqnarray}
  \label{eq:sLseq}
  f_{r's^2}^{(n')}(\lambda)
  =\frac{(-1)^{r's^2-1}}{2^{r's^2-4}}\sin^2\frac{\pi js}{s'}
  \sin^2\frac{\pi l}{s'}~, \quad
  (j=0,1,\cdots,r-1,~ l=0,\cdots,s'-1)
\end{eqnarray}
where $n'=n/s'$ and $r'=r/s'$ with $s'$ being the greatest common 
divisor of $r$ and $n$. 

\begin{figure}[h]
  \begin{center}
    \epsfig{file=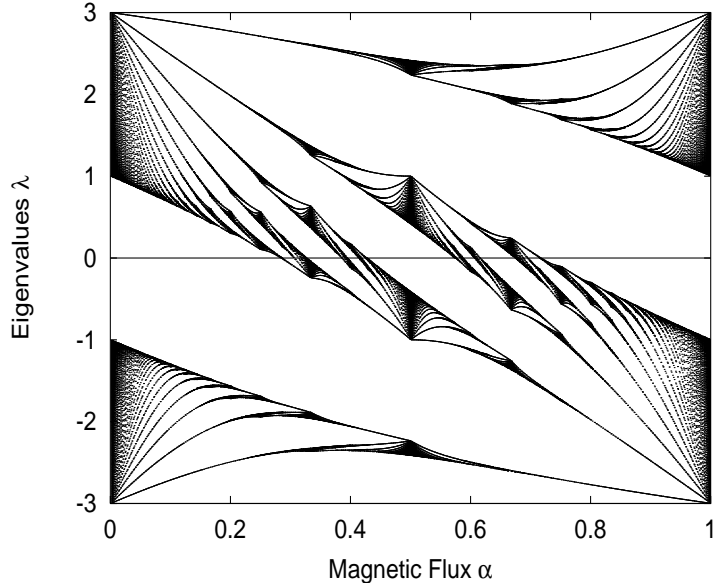,clip=,width=10cm,height=8cm,angle=0}
    \caption{The spectrum of $H$ for uniform magnetic fields 
      ($L=23$, $m=1$).}
    \label{fig:z23}
  \end{center}
\end{figure}

\begin{figure}[h]
  \begin{center}
    \epsfig{file=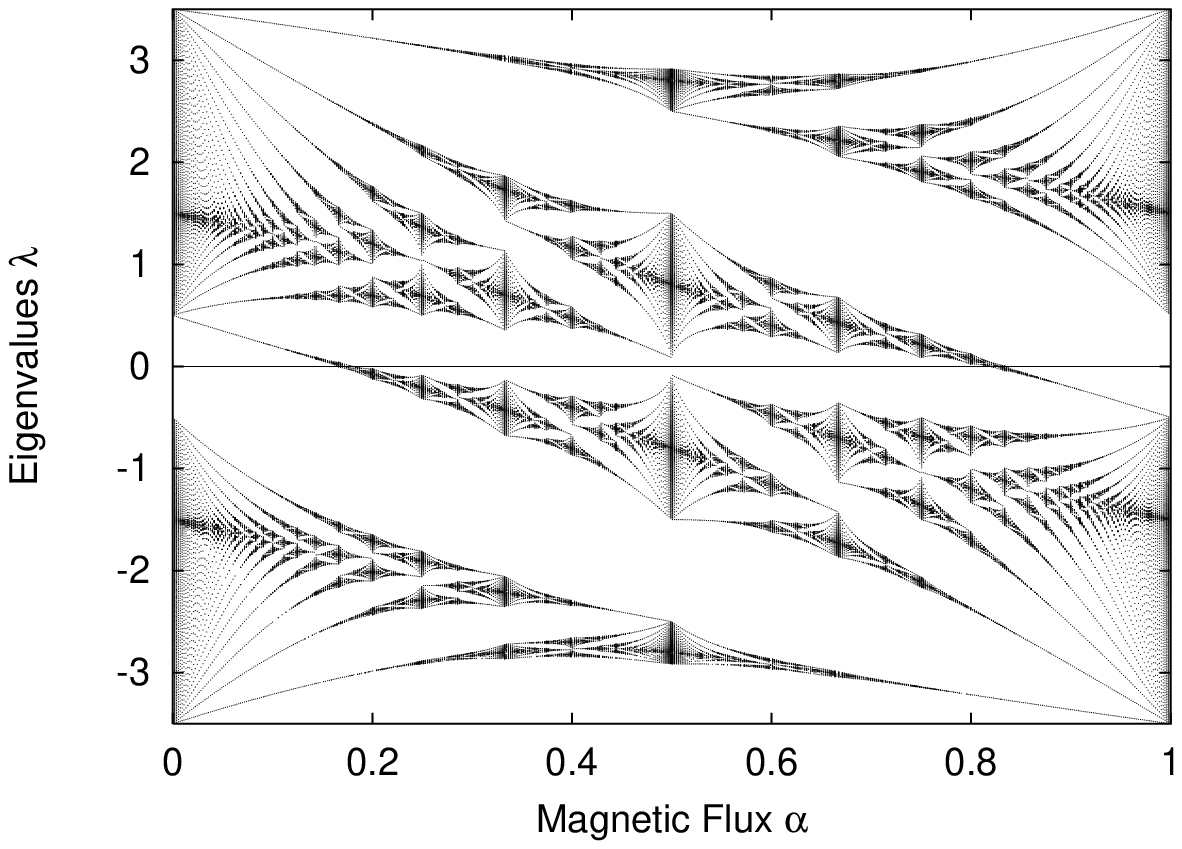,clip=,width=10cm,height=8cm,angle=0}
    \caption{The spectrum of $H$ for uniform magnetic fields 
      ($L=23$, $m=1/2$).}
    \label{fig:z23-0.5}
  \end{center}
\end{figure}

We see from (\ref{eq:sLseq}) that for each $j$ and $l$ there are 
$2r's^2$ distinct eigenvalues and they must lie one by one 
in the $2r's^2$ narrow intervals given by the inequality 
\begin{eqnarray}
  \label{eq:ieq}
  0\leq (-1)^{r's^2-1}2^{r's^2-4}f_{r's^2}^{(n')}(\lambda)\leq1~. 
\end{eqnarray}
Each interval contains exactly $rs'$ eigenvalues. Furthermore, 
the intervals themselves form roughly $2r's^2/n'$ clusters. 
This explains the characteristic features of the spectral flows 
of $H$ found by the numerical investigation \cite{fuji}. 
We can also count the multiplicity of the eigenvalues from 
(\ref{eq:sLseq}). In particular the multiplicity at $t=rn$ 
with $n$ and $r$ being coprime is exactly $r$. 
In Fig. \ref{fig:z23} the spectrum of $H$ for uniform 
magnetic field is shown. One easily see that the 
feather-shape gaps form a fractal pattern. It looks quite 
different from the butterfly diagram \cite{hof}. This is 
due the special choice of $m=1$ and the butterfly-like gaps 
appear for $m\ne1$. The spectrum for $m=1/2$ is shown in 
Fig. \ref{fig:z23-0.5} for comparison. 


In order to understand the appearance of the fractal structure of the 
spectrum we consider the infinite volume limit. The magnetic flux 
$\alpha$ defined by (\ref{eq:fs}) may be an arbitrary real 
number by considering the limit $t,L\rightarrow\infty$ with 
$\alpha=t/L^2$ fixed. In particular the eigenvalues at $\alpha=n/r$ 
with $r$ and $n$ being coprime positive integers form $2r$ bands 
given by the inequality 
\begin{eqnarray}
  \label{eq:rninf}
  0\leq(-1)^{r-1}2^{r-4}f_r^{(n)}(\lambda)\leq1~,
\end{eqnarray}
where $n$ specifies how these bands cluster each other. 
Roughly speaking, $n$ if $2n<r$ or $r-n$ if $2n>r$ stands 
for the number of near lying bands. An irrational flux is 
realized as an appropriate limit $r,n\rightarrow\infty$. This 
implies that the finite number of bands for a rational flux split 
into an infinite number of tiny bands (maybe a Cantor set). 
Though the spectrum appears smoothly varying with the flux 
due to the low resolution of the plot, such tremendous 
splittings and focusings of the bands take place 
continually during a small change of the flux. 

As was argued in Ref. \cite{fuji}, it is possible to find the 
index of the ODO $D$ defined by
\begin{eqnarray}
  \label{eq:NDO}
  D=1+\sigma_3\frac{H}{\sqrt{H^2}}~. 
\end{eqnarray}
The index of $D$ is given by 
\begin{eqnarray}
  \label{eq:indD}
  {\rm index}D={\rm Tr}\sigma_3\Biggl(1-\frac{1}{2}D\Biggr)
  =-\frac{1}{2}{\rm Tr}\frac{H}{\sqrt{H^2}}~,
\end{eqnarray}
where ${\rm Tr}$ implies the sum over the lattice coordinates as well as 
the trace over the spin indices. Since the trace on the RHS of this 
expression equal the number of positive eigenvalues of 
$H$ minus the number of negative eigenvalues, we can find 
${\rm index}D$ for constant field strength configurations by counting 
the root asymmetry of the secular equations (\ref{eq:sLseq}), where by 
root asymmetry of a polynomial equation we mean the number of positive 
roots minus the number of negative roots. In general the origin 
$\lambda=0$ lies out side of the intervals defined by the inequalities 
(\ref{eq:ieq}) as can be seen from Fig. \ref{fig:z23}. If this is 
satisfied, it is possible to relate ${\rm index}D$  with the root 
asymmetry of $f_{r's^2}^{(n')}(\lambda)=0$. We thus obtain for 
$t=nr=n'r's'{}^2$ 
\begin{eqnarray}
  \label{eq:index}
  {\rm index}D=-\frac{1}{2}rs'\sigma_{r's^2}^{(n')}~,
\end{eqnarray}
where $\sigma_{r}^{(n)}$ stands for the root asymmetry of 
$f_r^{(n)}(\lambda)=0$. 

\begin{figure}[h]
\begin{center}
    \epsfig{file=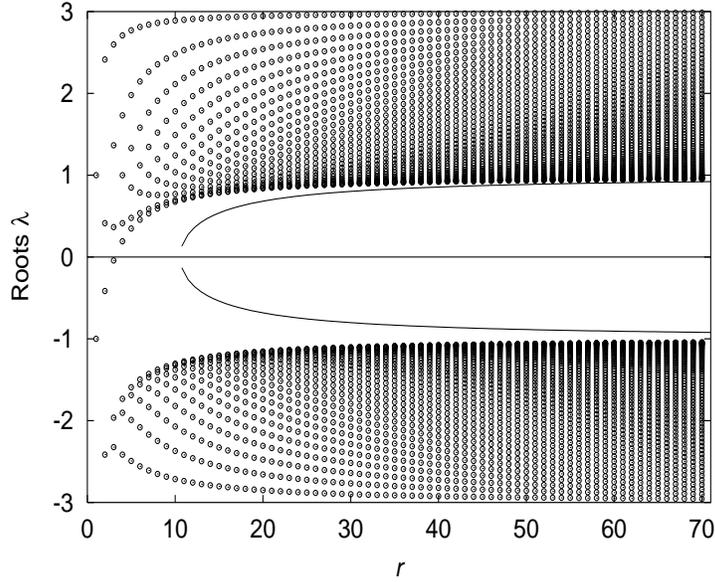,clip=,width=10cm, height=8cm,angle=0}
    \caption{Roots of  $f_r^{(1)}(\lambda)=0$ for $1\leq r\leq70$.
      The two curves indicate the known bound $|\lambda|\geq
      \sqrt{1-(2+\sqrt{2})\sin\frac{\pi}{r}}$.}
    \label{fig:roots_fr}
  \end{center}
\end{figure}

To understand $\sigma_r^{(n)}$ the roots of $f_r^{(1)}(\lambda)=0$ 
are plotted in Fig. \ref{fig:roots_fr}. As is easily seen, the root 
with minimum absolute value for each $r$ monotonically increases as 
$r$ and changes the sign from minus to plus at $r=4$. We thus find 
$\sigma_r^{(1)}=0$ for $r\leq3$ and $\sigma_r^{(1)}=2$ for $r\geq4$. 
In the case $n>1$ the behaviors of $\sigma_r^{(n)}$ are rather 
complicated for $r<4n$. 
However, we know $\sigma^{(n)}_r=0$ for $r=2n,~3n$ and 
$\sigma^{(n)}_r=2n$ for $r=4n,~5n,~\cdots$ from the factorization 
relation (\ref{eq:fp}). In fact it is possible to show $\sigma_r^{(n)}=2n$ 
for sufficiently large $r$ ($\geq4n$) rigorously by directly evaluating the 
spectral asymmetry of the hermitian matrix (\ref{eq:h}) for $p=q=0$ 
in the large $r$ limit \cite{kf}. We thus obtain for $r's^2\geq4n'$ or 
equivalently for $t=nr\leq L^2/4$ 
\begin{eqnarray}
  \label{eq:indth}
  {\rm index}D=-rs'n'=-rn=-\frac{1}{2\pi}\sum_{x}F_{12}(x)~,
\end{eqnarray}
where use has been made of (\ref{eq:fs}). This is the 
index theorem for the ODO (\ref{eq:NDO}) in the  
abelian gauge background in two dimensions \cite{NN2,fuji}. 

We have argued that the HWDO can be block-diagonalized for an arbitary 
uniform magnetic field and the spectrum is described by a relativistic 
analog of the Harper equation. We have found that the polynomials 
characterizing the spectrum possess a remarkable factorization property, 
from which we can understand the fractal nature of the 
spectrum and count the degeneracy of the eigenvalue. The root asymmetry 
of these polynomials turned out to be related with the index of the ODO. 
We have established the index theorem for the uniform magnetic field 
$|F_{12}(x)|\leq\pi/2$. The bound is of course not optimal and  
depends on the choice of the parameter $m$. 

\eject

\end{document}